\documentclass[conference]{IEEEtran}
\IEEEoverridecommandlockouts

\usepackage{cite}
\usepackage{amsmath,amssymb,amsfonts}
\usepackage{algorithmic}
\usepackage{graphicx}
\usepackage{textcomp}
\usepackage[algo2e]{algorithm2e} 
\usepackage{algorithmic}
\usepackage{algorithm}
\usepackage{xcolor}
\usepackage{listings}
\def\BibTeX{{\rm B\kern-.05em{\sc i\kern-.025em b}\kern-.08em
    T\kern-.1667em\lower.7ex\hbox{E}\kern-.125emX}}
\begin{document}

%\title{SchEdge: \underline{D}ynamic, \underline{A}daptive, and \underline{S}calable Scheduling Simulator for \underline{H}eterogeneous IoT}

\title{SchEdge: A Dynamic, Multi-agent, and Scalable Scheduling Simulator for IoT Edge}

\author{
\IEEEauthorblockN{Ali Hamedi, Amirali Ghaedi, Amin Soltanbeigi, Athena Abdi}
\IEEEauthorblockA{\textit{Faculty of Computer Engineering} \\
\textit{K.N.Toosi University of Technology}\\
Tehran, Iran \\
a\_abdi@kntu.ac.ir}
}

%\IEEEoverridecommandlockouts
%\IEEEpubid{\makebox[\columnwidth]{979-8-3503-7467-4/24/$\$$31.00 ©2024 IEEE \hfill}
%\hspace{\columnsep}\makebox[\columnwidth]{ }}

\maketitle

\begin{abstract}
This paper presents a dynamic, adaptive, and scalable framework for simulating task scheduling on the edge of the Internet of Things called "SchEdge". This simulator is designed to be highly configurable to reflect the detailed characteristics of real-world IoT. This framework focuses on online task scheduling and its multi-agent nature provides multiple schedulers to implement various scheduling schemes in parallel. SchEdge consists of two main parts the workflow and data flow. The workflow manages the schedulers' interaction with the application and environment while the data flow deals with the input application and its preprocessing. Combining these sections provides scalability, adaptability, and efficiency in the SchEdge. To validate the efficiency of this simulator, several experiments categorized as behavioral and technical analysis are performed to show its efficiency, scalability, and robustness. 

\end{abstract} 


\section{Introduction}

The Internet of Things (IoT) consists of interconnected devices that perform computations that are distributed. Along with the devices, distributed and centralized computations at the edge and cloud layers, improve the efficiency of these systems. The provided distributed computing at various levels makes them an appropriate option in many real-time applications including healthcare, industry, agriculture, smart grids, smart cities, and so on. The IoT systems are described in three layers the connected devices with limited computation capabilities, edge computing devices with high and low-latency computation capabilities, and the central cloud with the highest but latent computation capability.~\cite{b1,b2}. 

Due to IoT systems' dynamic and resource-constraint nature, executing tasks should be managed at various levels. Task scheduling performs this management by prioritizing available tasks and determining appropriate resources for execution. In this context, task scheduling optimizing the system’s constraints is an NP-hard problem that adapts varying workloads to the network conditions in real-time. This process dramatically affects the system’s performance and garnered significant attention from academia and industry, driving researchers to explore efficient methods. Due to the complexity of this problem, the proposed methods are approximate ones classified into heuristic, meta-heuristic, and learning-based schemes. Verifying the proposed methods and comparing them requires a reliable and comprehensive framework that reflects the target environment characteristics efficiently. ~\cite{b3,b4,b5}. 

The best verification framework is a real environment that shows all constraints and limitations. However, providing this real environment is very difficult and does not prepare repeatable conditions required for fair comparison due to its diversity. To address these challenges effectively, researchers rely on simulators to verify and compare their proposed schemes under controlled and repeatable conditions. In this context, the existence of a generally accepted framework that reflects all characteristics of the target environment and provides the possibility of implementing the various schemes is important. 
The available IoT task scheduling simulators often fail to meet the advanced research demands in this emerging trend. Key limitations of these simulators included computational inefficiency, difficulty in customization, and overly complex codebases. Moreover, they mostly rely on offline schemes and omit the online, live,  and iterative scheduling capabilities. Thus, verifying the dynamic and real-time approaches that require adaptive workflows, such as online schedulers and iterative decision-making strategies on these simulators is impossible~\cite{b6,b7,b8}.

In this paper, a lightweight and highly configurable simulator to emulate real-world IoT environments while maintaining simplicity and flexibility is designed. Our designed simulator called “SchEdge” leverages an encapsulated design for its components to be modular and comprehensive. SchEdge focuses on online task scheduling and provides a cycle-accurate policy for task execution through various scheduling schemes. This simulator is implemented in Python to provide a clear and extendable codebase, enabling users to adapt it to diverse research needs.
It accurately models IoT environments by simulating realistic data generation and task arrival processes, integrating an online scheduler to manage tasks at arrival. Additionally, the simulator tracks resource usage, monitors task execution, and captures system dynamics, enabling detailed evaluations of scheduling strategies and their impact on various aspects of performance such as makespan and energy consumption. Moreover, it provides a multi-agent and scalable framework that is well-suited for task scheduling in dynamic IoT environments. 

The remainder of this paper is organized as follows: Section 2 provides an overview of related work and their limitations. Section 3 details the methodology, describing the architecture, core components, and workflow of our designed simulator. Section 4 presents experimental results evaluating the simulator’s performance and adaptability. Finally, Section 5 discusses key insights and concludes the paper with potential future directions.

\section{Related Methods}
The increasing complexity and dynamic nature of IoT environments have created a growing demand for simulators that can support online scheduling strategies. Traditional simulators, while effective for static or semi-static scenarios, struggle to handle real-time task arrivals, dynamic workloads, and the heterogeneous nature of IoT devices. To address these challenges, online scheduling approaches have emerged as promising solutions. These methods focus on iterative decision-making, enabling systems to respond dynamically to changes in resource availability, energy constraints, and task priorities. Consequently, there is a critical need for simulation tools that can effectively model and evaluate these advanced scheduling strategies in realistic IoT environments.

EdgeCloudSim, an extension of CloudSim, introduces features such as task offloading, latency modeling, and scalability analysis, making it suitable for edge-specific environments. However, its architectural complexity and inability to support real-time scheduling limit its applicability for dynamic workloads. Similarly, iFogSim enables the modeling of fog and edge computing systems and supports features like latency analysis and resource allocation. Despite its strengths, iFogSim is limited by its reliance on static or semi-static scheduling approaches, and iFogSimit struggles to evaluate adaptive strategies required in evolving IoT systems. Both EdgeCloudSim and iFogSim face challenges in modeling dynamic environments where devices frequently join or leave the network, and workloads change continuously~\cite{b8,b9,b10}.

CloudSimPlus and GreenCloud excel in modeling static resource allocation for cloud-based, centralized systems with stable workloads and predefined task requirements. These simulators offer robust performance analysis but lack the flexibility to handle dynamic task generation, iterative scheduling, or real-time adaptability. Given the unpredictable and resource-constrained nature of modern IoT systems, they fail to evaluate strategies for dynamic environments where adaptability and responsiveness are essential~\cite{b11,b12}.

In summary, while existing simulators like EdgeCloudSim and iFogSim offer valuable capabilities for modeling edge and fog computing systems—such as task offloading, latency analysis, and resource allocation—they rely heavily on static or semi-static scheduling approaches. This restricts their effectiveness in dynamic IoT environments, where devices frequently join and leave, and workloads continuously evolve. Similarly, cloud-based simulators like CloudSimPlus and GreenCloud perform well in static resource allocation and performance analysis for centralized systems with stable workloads but lack critical features such as real-time adaptability, dynamic task generation, and iterative scheduling. These limitations hinder their ability to provide real-world-like simulations necessary for evaluating advanced scheduling strategies in dynamic and heterogeneous IoT systems.

\section{Details of the designed SchEdge}
\subsection{Simulator Overview}
To support IoT task scheduling research, we developed a simulator that combines computational efficiency, modularity, and real-world representation. This tool is specifically designed to address the challenges of IoT scheduling, including dynamic workloads, resource constraints, and the need for real-time adaptability. Many existing simulators fail to meet the demands of iterative, online approaches, often lacking the flexibility and efficiency required for such tasks.

The simulator is guided by three core principles: modularity, to ensure customization and scalability; lightweight design, to optimize computational efficiency for resource-constrained systems; and realism, to closely emulate IoT environments and their inherent complexities. Together, these principles create a high-performing platform for IoT research.

Key features include a modular architecture for seamless experimentation, realistic workload generation to simulate dynamic IoT conditions, and real-time schedulers for managing tasks as they arrive. Additionally, the simulator incorporates comprehensive performance tracking, capturing metrics such as resource utilization, system dynamics, and task execution efficiency in terms of makespan and energy consumption. By aligning with these design principles, the simulator provides a robust and adaptable foundation for evaluating and improving scheduling mechanisms in dynamic IoT environments.

\subsection{Architecture Description}
The architecture of the designed SchEdge consists of two main sections: the workflow layer that manages the interaction of the scheduler, environment, and tasks along with the dataflow layer that handles the generation, management, and preprocessing of incoming applications. Together, these sub-systems provide scalability, adaptability, and efficiency that make the designed simulator an effective framework for IoT task scheduling research.

The workflow section consists of several key components that manage the simulator’s core processes and ensure efficient task scheduling. These components are the environment, scheduler, and state of the system. The environment is the foundational unit coordinating the overall function of the simulation. It specifies the essential setup for the simulation, manages dynamic configurations, such as adding or removing devices, and manages the interface between the workflow and dataflow sections. The main simulation loop, state updates, and system responses are all organized by the environment. Through the modularization of underlying operations, the environment provides a robust and flexible framework for real-time experimentation.
The scheduler is a key component responsible for task allocation and prioritization. It enables interaction with the environment to allocate tasks to processing elements and collect feedback, ensuring dynamic and adaptive decision-making. Designed to be highly configurable, the scheduler supports a range of online and adaptive approaches, including heuristic, metaheuristic, and learning-based methods.

The state is the most critical component of the simulator’s architecture, serving as the single source of truth for the system’s current conditions. While demonstrating the central bridge between the environment and the scheduler, the state ensures robust interactions and consistent decision-making. It synchronizes all live data to provide a compact yet comprehensive representation of the simulation environment, including application statuses, device statuses, and scheduling conditions.
The application status representation within the state presents application progress through a detailed categorization of tasks. This includes “Remaining-Tasks”, which are tasks that have arrived from the dataflow layer but are not yet executing; “Running-Tasks”, which represent tasks currently in execution; and “Finished-tasks”, which catalog tasks that have been completed. 
Once all tasks for an application are marked as finished, the job is removed from the state, but its record remains accessible for tracking and validation purposes. 

Device statuses include core availability, queue capacity, and battery levels. The simulator models devices heterogeneously, with cores having distinct specifications and individual queues for task allocation. Edge devices' battery levels are also tracked, impacting scheduling and aligning with real-world IoT requirements.
By centralizing these metrics, the state enables consistent observation, informed decision-making, and effective real-time management, making it indispensable to the simulator. %Fig.~\ref{fig:arch} shows the workflow's components and their relation. 

%\begin{figure}
%    \centering
%    \includegraphics[width=0.8\linewidth]{Architecture.png}
 %   \caption{Components of the workflow section of our proposed SchEdge}
%    \label{fig:enter-label}
%\end{figure}

While the workflow layer manages the core interactions and decision-making processes within the simulation, it relies on the dataflow layer to supply the dynamic inputs needed for realistic and adaptive experimentation. The collaboration between these layers ensures a seamless integration of task scheduling, system monitoring, and real-world representation.
The dataflow section provides the foundational inputs required for the simulator, including dynamic task generation and preprocessing. It includes specialized modules that emulate realistic IoT environments by generating, processing, and delivering data efficiently to support the task scheduling process. 

The simulator’s dataflow section includes two parts: the data generator and the task manager. 
The data generator is responsible for creating the foundational entities of the simulation, including applications, tasks, and devices. Designed for flexibility and adaptability, it allows users to dynamically configure a wide range of parameters, ensuring the simulation can replicate diverse IoT research scenarios.
Applications are defined with varying sizes and deadlines, representing a spectrum of application complexities. Tasks within the applications are adjustable in terms of computational load, safety requirements, dependencies, and arrival rates. Device specifications are equally flexible, supporting heterogeneous setups where processing elements differ in capabilities, core counts, and resource constraints like queue capacities and battery levels. This heterogeneity enables the accurate modeling of diverse IoT environments, from small-scale systems to large, resource-constrained edge networks.
The dynamic and configurable design of the data generator empowers researchers to create tailored experimental setups, ranging from uniform workloads to highly complex, realistic scenarios, making it indispensable for modeling complex IoT workflows.

\begin{figure}
    \centering
    \includegraphics[width=0.8\linewidth]{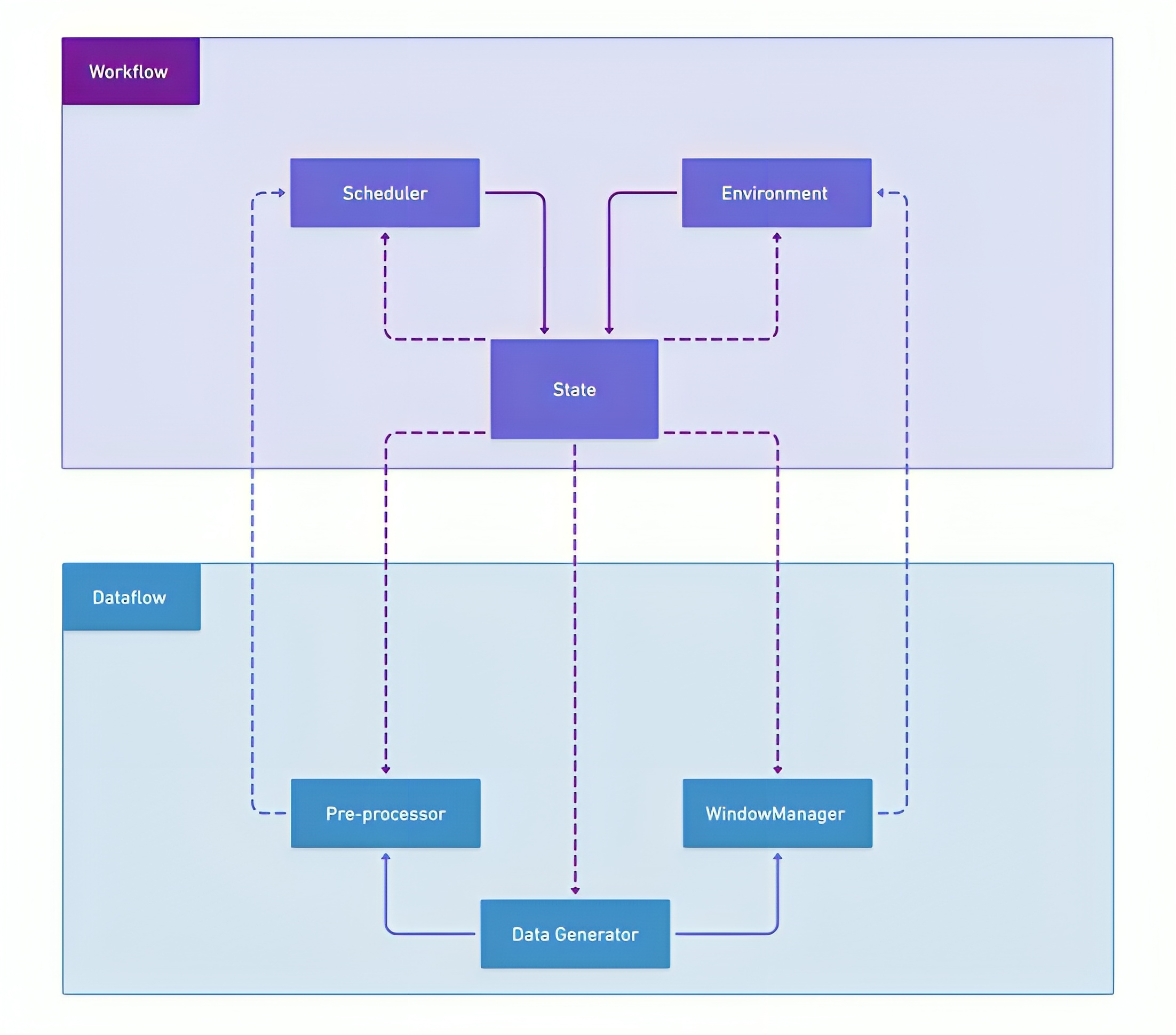}
    \caption{The detailed architecture of the proposed SchEdge}
    \label{fig:wd}
\end{figure}

The Task Manager ensures seamless integration between the generated data entities by the data generation component and the environment by managing how tasks are arrived at in the environment and prepared for execution. This is achieved through two submodules, the "WindowManager" and the "Preprocessor".  The Window Manager retrieves data from the database and delivers it to the environment in configurable units, ensuring smooth and controlled task flow. It selects a window of tasks from a defined number of jobs, shuffles them, and delivers them to the environment at specified cycle intervals. 
The Preprocessor prepares tasks for scheduling by filtering out incomplete tasks that are not yet ready for execution. Then it prioritizes the ready tasks by sorting them based on different criteria like the number of successors in the application graph, ensuring that critical tasks are addressed first.
By combining these components, the dataflow system provides a strong foundation for the simulator, enabling realistic modeling of IoT environments and adaptable task scheduling workflows. The modular design ensures flexibility, allowing researchers to replicate a wide variety of IoT scenarios and evaluate diverse scheduling strategies. Fig.~\ref{fig:wd} shows the architecture of the proposed SchEdge considering its workflow and dataflow sections. 

\subsection{Simulation Process}
The simulation process is structured into a series of iterative cycles, beginning with an initialization phase and followed by a dynamic main loop that coordinates task scheduling, state updates, and performance monitoring. This iterative workflow ensures the simulator can accurately model the real-time and adaptive nature of IoT systems.
The simulation begins with the setup of the environment, which acts as the central unit of the simulation. This setup triggers the data generation process, creating the foundational entities required for the simulation, including applications, tasks, and devices. Each entity is generated dynamically based on user-defined configurations, allowing for flexibility in attributes such as task dependencies, priorities, and device capabilities.
Once the data generation is complete, essential components, including the scheduler and dataflow modules, are initialized. 

The scheduler is configured for task allocation and iterative updates, while the dataflow modules, such as the Window Manager and Preprocessor, are prepared to manage task arrival and processing. The state is then populated with the initial statuses of tasks, jobs, and devices, ensuring that the system is ready for dynamic operations.
With all components initialized, the simulation transitions into the main loop, where iterative cycles drive task scheduling, state updates, and performance monitoring.
After initialization, the simulator configures the connections between its core components, including the environment, scheduler, and dataflow modules. This phase ensures that communication pathways are established, allowing seamless interaction and synchronization during the simulation. 

Following initialization, the simulation progresses iteratively through a series of cycles, each consisting of three main stages: task scheduling, state updates, and monitoring. In the scheduling step, the scheduler allocates tasks to devices based on the current state. Tasks are prioritized and matched to devices with sufficient capacity, considering attributes such as queue availability and battery levels. Once tasks are scheduled, the environment updates the state to reflect execution progress. 
New tasks are periodically introduced into the simulation via the Window Manager, simulating real-time or batch task arrival patterns. These tasks are processed by the Preprocessor, which ensures they are ready for scheduling by verifying dependencies and organizing them based on priority criteria, such as the number of successors in the application graph. Monitoring is performed periodically after a configurable number of cycles, capturing critical system data to evaluate the simulation’s progression.

Performance metrics play a key role in understanding the simulator’s behavior, focusing on both the performance of the scheduling agents and the system’s operational efficiency. Metrics such as reward signals, loss values, and task scheduling objectives provide insights into how well the agents achieve their goals. At the system level, metrics like cycle execution time and average memory usage per process highlight the efficiency and stability of the simulation. These measures ensure the simulator’s capability to operate effectively under varying conditions and support its use in evaluating scheduling strategies.

A critical aspect of the simulation workflow is maintaining accurate and synchronized updates of the system’s current conditions. This is achieved through the state update process, which serves as the backbone of the simulator’s iterative operations. Managed by the environment, the state update process ensures that both device and job statuses are accurately reflected, enabling seamless interaction among all components. Device updates focus on tracking the execution of tasks and managing resource consumption. During each cycle, the environment reduces the execution time of running tasks, identifies completed tasks, and removes them from the state. 

\begin{figure}
    \centering
    \includegraphics[width=0.8\linewidth]{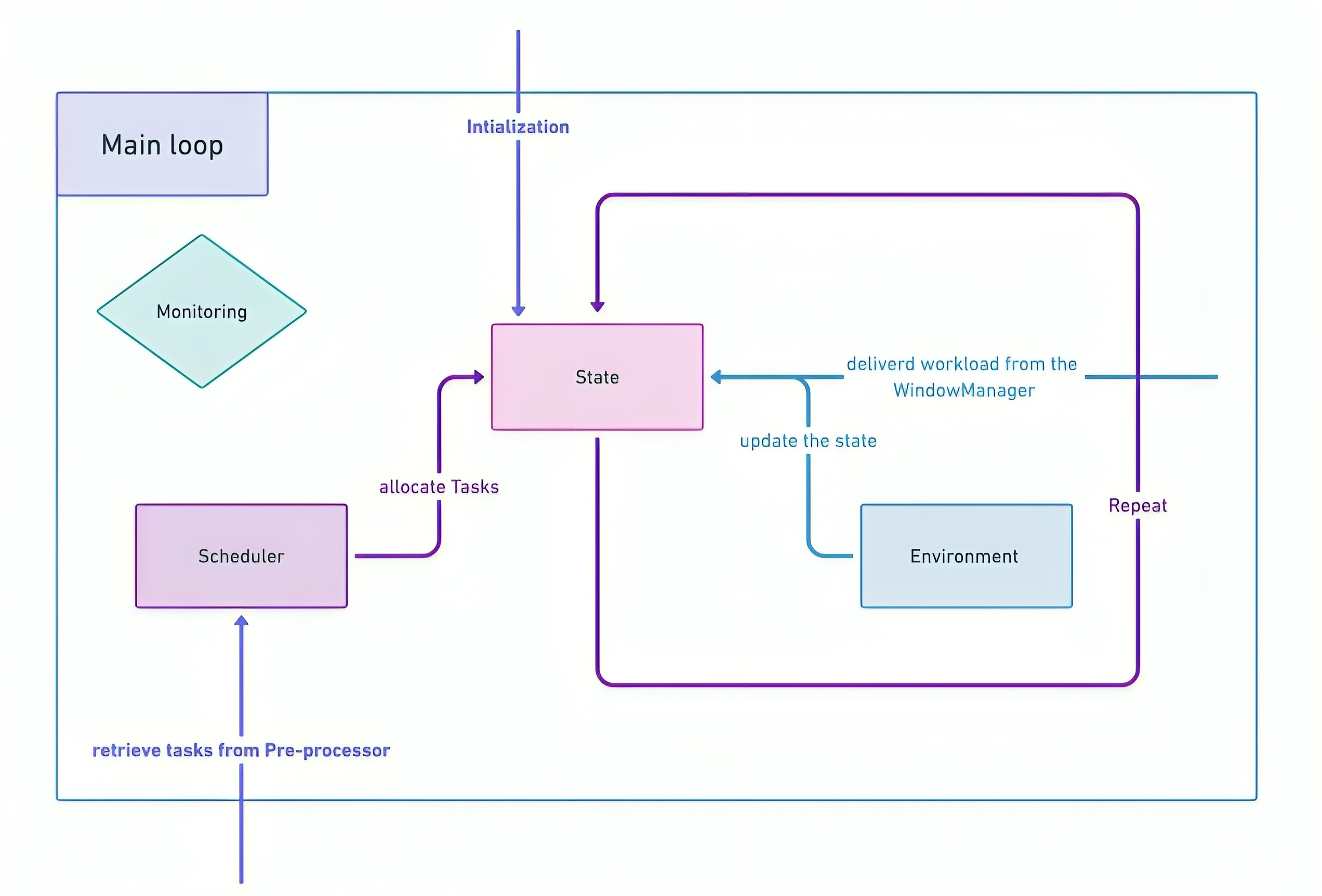}
    \caption{The simulation process details of the proposed SchEdge}
    \label{fig:sim}
\end{figure}

Application updates are managed collaboratively by the dataflow components and the environment. The window manager introduces tasks into the simulation, which are added to the state for processing. As tasks progress through execution stages, the environment tracks their status and transitions completed tasks into a finalized state. When all tasks within an application are completed, the environment marks it as finished, ensuring that job progress is consistently and accurately represented in the state. This dynamic update process supports real-time synchronization across components and maintains a clear overview of system conditions.

The scheduler and the preprocessor collaborate to ensure efficient task execution. The preprocessor retrieves tasks from the state, verifies their readiness by checking dependency constraints, and organizes them based on predefined criteria, such as the number of successors in their directed acyclic graphs (DAGs). It then presents the prioritized tasks to the scheduler. The scheduler, using this structured input, allocates tasks to devices based on resource availability and scheduling objectives. The details of the simulation process of the proposed simulator are summarized in Fig.~\ref{fig:sim}.

\section{Experimental Results}

To evaluate the simulator’s capabilities in handling dynamic task scheduling scenarios under realistic and flexible conditions several experiments are performed. These experiments investigate the simulator’s ability to manage diverse workloads, optimize resource utilization, and maintain runtime stability while demonstrating its support for adaptive scheduling mechanisms.

\begin{figure*}
    \centering
    \includegraphics[width=1\linewidth]{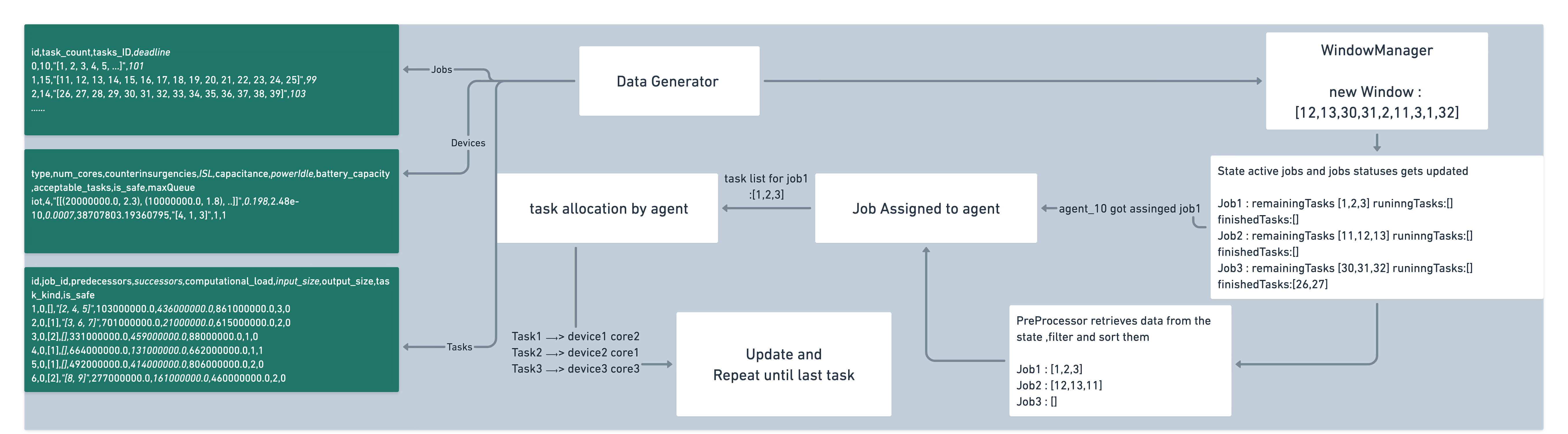}
    \caption{Behavioral analysis of the proposed framework on a case study of three applications}
    \label{fig:out}
\end{figure*}

\subsection{Setup}
The simulator was configured to emulate realistic IoT environments, with dynamic workloads and a heterogeneous device setup. A total of 10,000 applications each consisting of 20 to 60 tasks represented as directed acyclic graphs (DAGs) are generated. These tasks were characterized by attributes such as computational load, input/output sizes, and safety requirements, ranging from 1 MB to 1 GB. Dependencies were dynamically assigned to reflect real-world precedence constraints, ensuring the workload closely mirrored the complexity of IoT systems.

The infrastructure was composed of IoT, MEC, and Cloud devices to capture the diversity of IoT processing nodes. IoT devices, numbering 100, were modeled with 4,8, or 16 cores, small queue capacities, and battery levels ranging from 36 to 41 Wh, simulating resource-constrained environments. MEC devices were configured with 50 units, featuring 16,32, or 64 cores and higher power consumption, representing intermediate processing nodes. The Cloud server, with its infinite cores and absence of battery constraints, served as the high-performance backend for computationally intensive workloads.

Schedulers were implemented in a multi-agent setup using the sample configuration as the Asynchronous Advantage Actor-Critic (A3C) algorithm. The simulation supported 24 parallel agents operating over 10,000 iterations, dynamically adjusting task arrivals through a configurable window size of 1,000 tasks and up to 40 applications. These configurations provided a flexible and robust framework to evaluate adaptive scheduling strategies under diverse target scenarios.

\subsection{Efficiency of SchEdge in a case study}
To demonstrate the efficiency of our SchEdge, two primary observations from the simulation are studied: analysis of addressing behavioral challenges in IoT scheduling and evaluation of its technical efficiency in terms of resource utilization and runtime stability. 
The behavioral analysis focuses on the role and effectiveness of individual simulator components, while the technical one highlights the simulator’s ability to maintain efficient resource management and stable performance. Last the features and characteristics of our proposed SchEdge are compared to related simulators. 

\subsubsection{Behavioral Analysis}

In the first phase, the simulator performs Initialization and data generation. During this phase, a time complexity proportional to the input's DAG's size and dependency density is generated. Afterward, the data was stored in .csv files, categorized as applications and tasks. Analysis of the stored data confirmed a balanced distribution of task attributes, including size, priority, and dependencies, providing a strong foundation for diverse scheduling scenarios.

Next, the WindowManager ensured a smooth and manageable workload throughout the simulation by dynamically delivering tasks. Every 15 cycles, it provided a configurable window of up to 1,000 tasks from a maximum of 40 applications. This approach introduced flexibility while maintaining a controlled task flow, enabling an effective evaluation of adaptive scheduling strategies under realistic workloads.
Leveraging the state-centric architecture, in each simulation cycle, agents were assigned applications and obtained the relevant tasks from the Preprocessor component. The Preprocessor filtered tasks based on their readiness and sorted them according to priority, presenting the tasks in an optimized order. Agents iteratively attempted to schedule tasks, which were allocated to devices based on the model’s scheduling policy. This policy prioritized predefined constraints such as queue capacity while optimizing for execution time, energy consumption, safety compliance, and task compatibility. Tasks that failed to meet these constraints remained unallocated 
and carried forward for rescheduling in future iterations. Over successive iterations, agents demonstrated improved scheduling accuracy, as the number of successful task allocations per cycle increased as a result of the agents’ learning to avoid earlier mistakes. Fig.~\ref{fig:out} shows the behavioral analysis of our proposed framework on a sample stream of three applications. 

The simulator’s scalability was tested by dynamically adding or removing devices during the simulation. Each iteration introduced a small probability (0.005–0.01\%) of either adding or removing a device. To avoid significant imbalances, no more than three consecutive additions or removals were allowed. New devices adhered to the predefined proportions: 33\% MEC devices and 66\% IoT devices. These changes were reflected across the environment, the live state, and the agents’ action space. After 10,000 simulation steps, approximately half of the original environment had changed, yet agents maintained stable learning performance, highlighting the simulator’s adaptability to evolving conditions.

\subsubsection{Technical Analysis}
\begin{figure}
    \centering
    \includegraphics[width=0.8\linewidth]{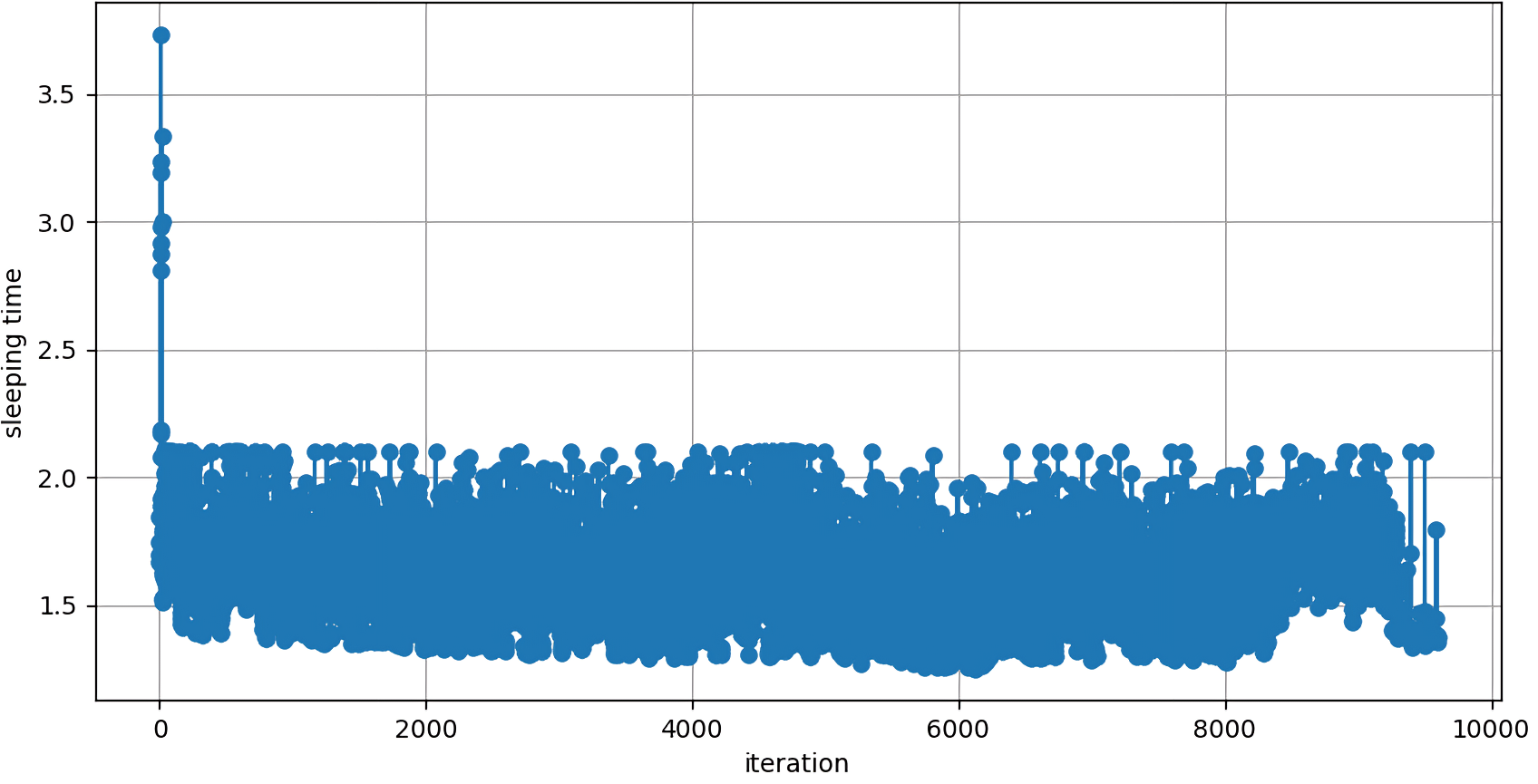}
    \caption{Stability of the iteration times over the various simulation cycles}
    \label{fig:exp1}
\end{figure}

Despite the computational demands of the multi-agent setup utilizing PPO and A3C algorithms—a challenging scenario for resource management—the simulator maintained stable and efficient resource consumption throughout the process. These results emphasize the simulator’s capability to manage high-demand workloads effectively while supporting scalable IoT scheduling applications. Fig~\ref{fig:exp1} shows the iteration times of the simulation cycles. 

As this figure shows, the simulator demonstrated stable and efficient iteration times across all simulation cycles, validating its reliability and adaptability for dynamic IoT environments. At the start, iteration times were elevated due to the initialization overhead of tasks and processes. However, the times quickly stabilized within a consistent range as the simulation progressed. Importantly, iteration times remained bounded and did not grow exponentially over prolonged runs, even with variations caused by task arrivals and processing complexities. This stability ensures the simulator can execute long-term simulations reliably without performance degradation.

\begin{figure}
    \centering
    \includegraphics[width=0.6\linewidth]{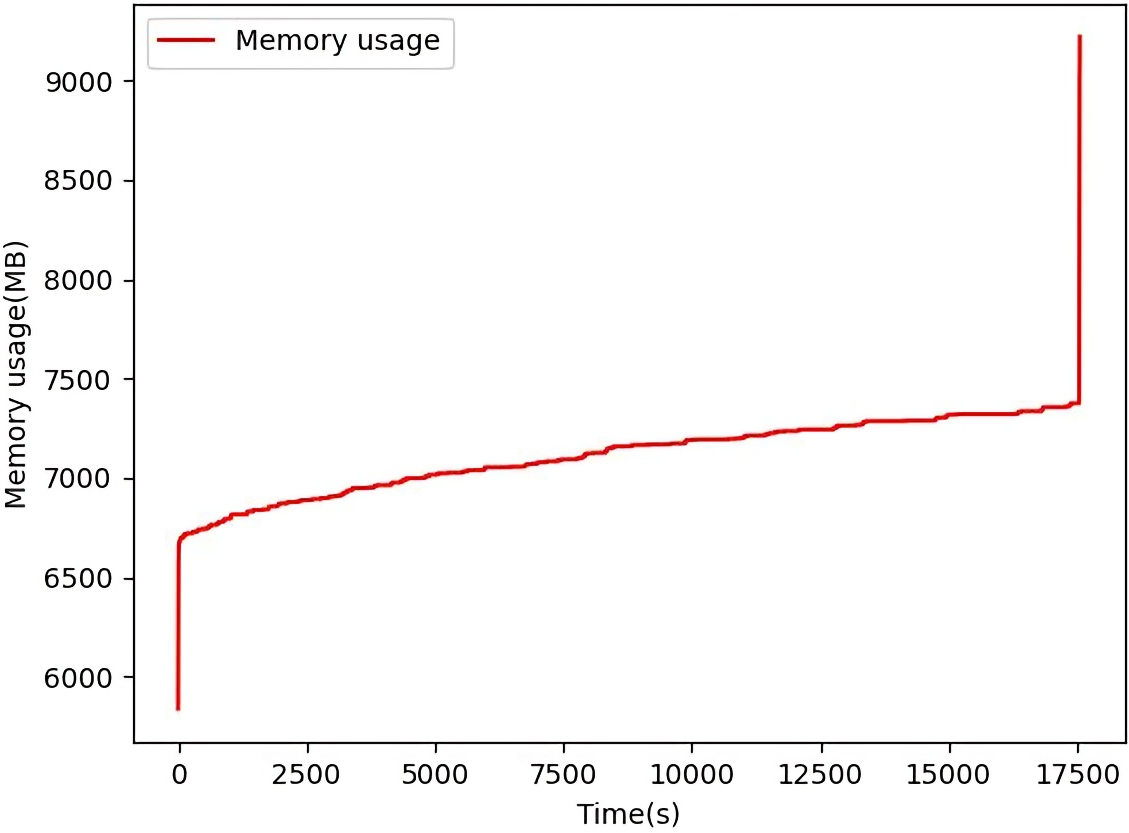}
    \caption{The total memory usage of the simulator during its operation}
    \label{fig:exp2}
\end{figure}

Memory usage also remained steady throughout the simulation cycles, with minor increases observed only during the initialization phase and the finalization phase as processes were being set up and closed. The profiler sampled memory usage every 10 seconds, capturing the consistent behavior of resource consumption throughout the simulation. Fig.~\ref{fig:exp2} shows the total memory usage of the simulator during its operation. 

As this figure shows, beyond the initialization and finalization phases, memory usage exhibited stable trends, regardless of workload complexity or agent count. The increase in initial and final memory usage is due to preparation and saving the results which are negligible in time. This efficient and predictable resource utilization makes the simulator appropriate for larger IoT systems. Together, these findings demonstrate the simulator’s ability to maintain consistent performance under diverse and dynamic conditions, supporting long-term, adaptive experimentation in IoT environments. 

\subsubsection{Comparing SchEdge to other simulators}
To further assess the effectiveness of SchEdge in handling the scheduling of IoT tasks, we compare its capabilities with existing simulation tools, including iFogSim, EdgeCloudSim, and CloudSimPlus~\cite{b8,b9,b10}. 
Table~\ref{table:comp} shows the comparison based on key features essential for dynamic and scalable scheduling in heterogeneous IoT environments. A key distinction is that SchEdge is implemented in Python, whereas the other simulators are written in Java. While Java-based simulators may offer faster execution, SchEdge’s Python-based design provides a significant advantage—it seamlessly integrates with the latest machine learning and optimization libraries, allowing for rapid experimentation and customization. Another major advantage of SchEdge is its built-in support for both online and offline scheduling schemes, unlike iFogSim, EdgeCloudSim, and CloudSimPlus, which do not support online scheduling in their basic versions. 

\begin{table*}[]
    \centering
    \caption{Comparison of the proposed SchEdge to related simulators in terms of their key features}
    \label{table:comp}
    \begin{tabular}{|c|c|c|c|c|}
    \hline
    \textbf{Feature} & \textbf{SchEdge} & \textbf{iFogSim} & \textbf{EdgeCloudSim} & \textbf{CloudSimPlus}\\
    \hline
    \hline
    Fault Handling (Reactive / Proactive ) &
    Yes  / Yes &
    No / No &
    No / No &
    No / No \\
    \hline
    Custom Scheduling Policies &
    online and offline learning &
    Offline learning &
    Offline learning &
    Offline learning \\
    \hline
    Energy Efficiency Consideration &
    Limited & Limited & Limited & No \\
    \hline
    Effect of Increasing Tasks on Performance &
    Scales efficiently &    Performance Degrades &
    Performance Degrades &    Struggles at 10,000+ tasks \\
    \hline
    Edge Layer Support &    Yes&    Yes&    Yes&    No \\
    \hline
    Data Transfers and Network Features & 
    Limited & Moderate & Detailed & Basic \\
    \hline
    Flexibility in Defining Custom Topologies &
    High & Limited & High & Not supported \\
    \hline
    Dynamic Device Join/Leave & Supported &  Limited & 
      Supported & Not supported \\
      \hline
      Scheduling Adaptability & Yes & Fixed &
    Yes & Fixed \\
\hline
Network Topology Adjustments & Completely Dynamic &
Static & Dynamic & Static \\
\hline
    \end{tabular}
\end{table*}

Although some studies have integrated reinforcement learning-based methods into iFogSim and EdgeCloudSim, these implementations required significant modifications to the source code, making them less accessible for adaptive scheduling research. When considering performance scalability, SchEdge is designed to scale efficiently as the number of tasks increases. In contrast, iFogSim and EdgeCloudSim experience performance degradation, while CloudSimPlus struggles when handling more than 10,000 tasks. This highlights SchEdge’s robustness in managing large-scale IoT workloads. In terms of network features and data transfers, SchEdge provides basic task-related network modeling—less detailed than some alternatives but sufficient for scheduling-focused studies. Meanwhile, flexibility in defining custom topologies is another area where SchEdge excels. Unlike iFogSim, which requires modifications to support non-hierarchical networks, and CloudSimPlus, which lacks support for custom topologies altogether, SchEdge supports any user-defined topology. This allows researchers to model arbitrary network configurations that more accurately reflect real-world IoT deployments. 

\subsubsection{Scalability Analysis of SchEdge}
Scalability is a crucial requirement in IoT environments, where the number of devices can dynamically increase or decrease over time. An effective IoT simulator must be able to model this dynamism by adapting to device mobility, scheduling flexibility, and network topology changes while maintaining performance and stability. One of SchEdge’s key advantages is its flexible device churn simulation, which allows for both user-defined control and probabilistic modeling. 
Users can manually specify when and how devices join or leave the network, providing a structured way to test specific scenarios and conduct controlled experiments.

The scalability capability of SchEdge in terms of three main parameters: dynamic device join/leave, scheduling adaptability, and network topology adjustment is verified in the last three rows of table~\ref{table:comp}. 
Unlike CloudSim and CloudSimPlus, which are designed for static environments, SchEdge and EdgeCloudSim enable dynamic device interaction. However, while EdgeCloudSim allows device movement, it does not offer full support for arbitrary topology changes as SchEdge. Moreover, iFogSim’s original version lacks proper support for device churn and mobility, requiring significant modifications to accommodate dynamic scheduling and topology adjustments. The updated iFogSim2 version improves on this by supporting dynamic scheduling and network changes but still lacks native mobility modeling.

\begin{figure}
    \centering
    \includegraphics[width=0.6\linewidth]{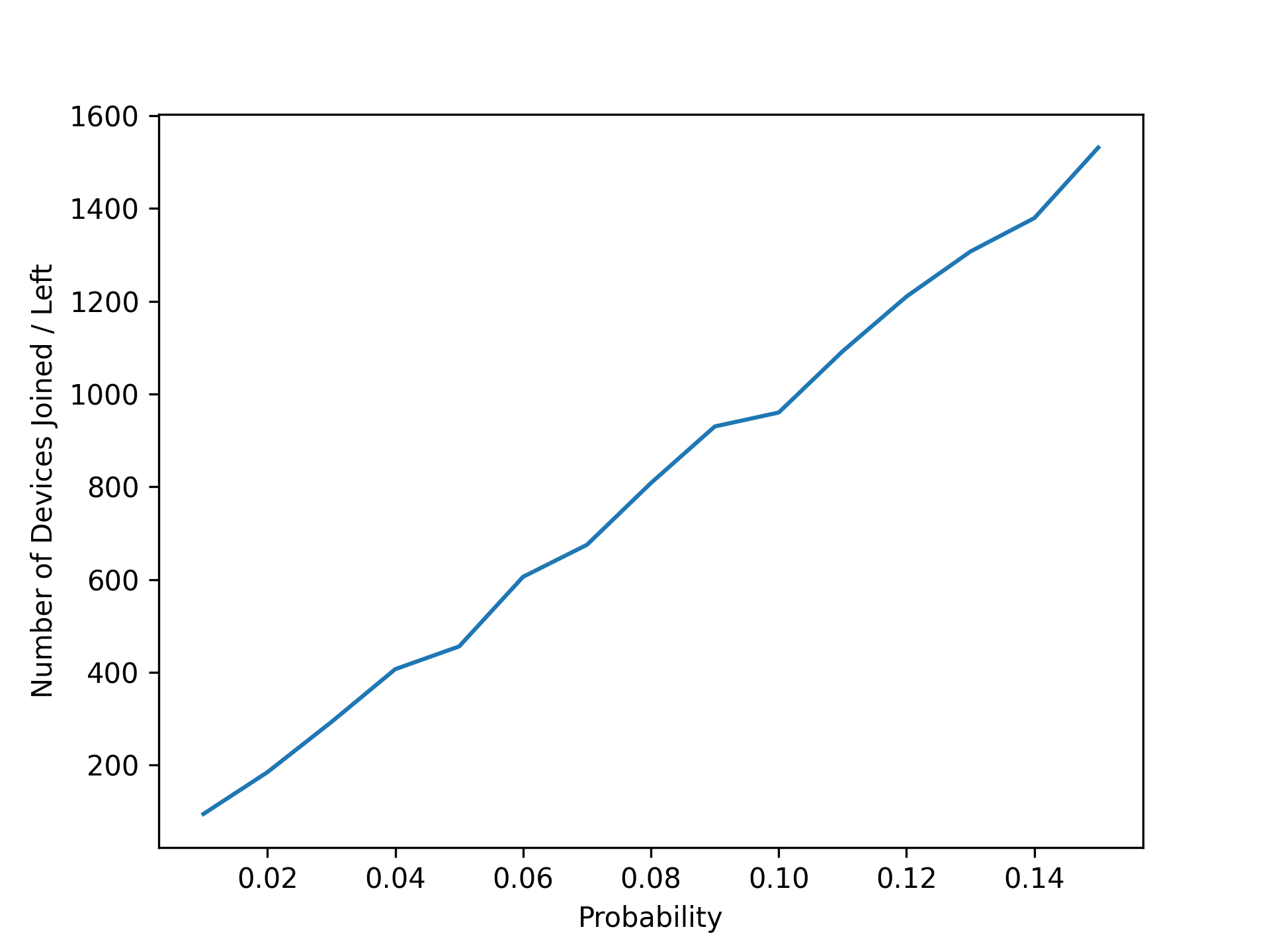}
    \caption{Capability of SchEdge in handling network changes}
    \label{fig:sca}
\end{figure}

To further analyze the ability of SchEdge to simulate network changes, Fig.~\ref{fig:sca} presents the average number of devices added and removed under different probability configurations ranging from 0.01 to 0.15. These results shows how the simulator dynamically adjusts the network over time, providing a reliable platform for evaluating scheduling strategies in unpredictable conditions.

\section{Conclusion}
In this work, a lightweight and configurable simulator called "SchEdge" that emulates real-world IoT environments for task scheduling research is presented. This simulator has a modular architecture, dynamic dataflow systems, and robust workflows. It supports diverse scheduling schemes including reinforcement learning (RL)-based methods. 
Experimental results demonstrated its stability, memory efficiency, and adaptability to dynamic workloads, while its Python-based implementation ensures accessibility and ease of customization. 
Due to the mentioned characteristics, the proposed simulator addresses critical gaps in existing tools, offering a robust, scalable, and flexible platform for IoT scheduling research. By supporting iterative, real-time experimentation, it provides a foundation for advancing scheduling strategies in dynamic and resource-constrained environments.
Future work will focus on expanding the simulator’s scalability, incorporating real-world datasets, and exploring additional scheduling paradigms, such as latency-aware or predictive methods. Further developments may include interfacing with real IoT devices and extending support to hybrid edge-cloud systems.

\textbf{AI Usage Statement}

The authors confirm that no artificial intelligence tools were used in the preparation of this article.

\end{document}